\documentstyle[]{article}
\begin{document}
 \title{Influence of Gravitation on Mass-Energy Equivalence Relation}
\author{R.~V.~R.~Pandya\thanks{Email address: {\sl
rvrpturb@uprm.edu}}, \\Department of Mechanical Engineering,\\
University of Puerto Rico at Mayaguez, Mayaguez,
PR 00680, USA }
\date{\today}
\maketitle

\begin{abstract}

We study influence of gravitational field on the mass-energy
equivalence relation by incorporating gravitation in the physical
situation considered by Einstein (Ann. Physik, 17, 1905, English
translation in ref. [1]) for his first derivation of mass-energy
equivalence. In doing so, we also refine Einstein's expression
(Ann. Physik, 35, 1911, English translation in ref. [3]) for
increase in gravitational mass of the body when it absorbs $E$
amount of radiation energy.
\end{abstract}
\section{Introduction}
Different forms of mass-energy equivalence relation existed even
before Einstein's first derivation of the relation
\cite{Einstein05} and which have been reviewed along with other
developments on the relation after the year 1905 (see Ref.
\cite{Fadner88} and references cited therein). These relations
were obtained in the absence of gravitational field. In this
paper, we show that the presence of gravitational field affects the
mass-energy equivalence relation. In doing so, we include
gravitational field in the physical situation considered by Einstein
\cite{Einstein05} and derive the general mass-energy equivalence
relation which reduces to the well known Einstein's relation in
the absence of gravitation. We also suggest refinement to the
expression, provided by Einstein \cite{Einstein11}, for increase
in gravitational mass due to the absorption of radiation energy
$E$.

\section{Analysis} Consider a `stationary' reference frame $S_s$
with coordinate axes $x,y,z$ and another moving reference frame
$S_v$ having coordinate axes $x',y',z'$ which are parallel to axes
$x,y,z$, respectively. Let $S_v$ is moving along the $x$ axis with
a constant translational velocity $v$ as measured in $S_s$.  At
time $t=0$ measured in $S_s$, let the origins of $S_v$ and $S_s$
coincide. Also, consider these two reference frames in the
homogeneous gravitational field whose potential is $\Phi_s$ and
$\Phi_v$ as measured in $S_s$ and $S_v$, respectively. For the
homogeneous gravitational field with acceleration of gravity
$\gamma$ and lines of force of the gravitational field in the
negative direction of the axes $z$ and $z'$, we have
$\Phi_s=\Phi_v=\gamma h$. Here $h$ is distance along the $z$ and
$z'$ axis from the origins of $S_s$ and $S_v$, respectively.

Consider a body $W$ of mass $M_s$ kept at rest in $S_s$ on the $z$
axis at a distance $h$ from the origin, i.e having coordinates
$(0,0,h)$. At some instance it emits in two opposite directions
equal quantity of light having energy $L/2$ where $M_s$ and $L$
are measured in $S_s$. The conservation of energy principle for
this situation in $S_s$ can be written as
\begin{equation}
E_0=E_1+L\frac{\Bigl(1+\frac{\Phi_s}{c^2}\Bigr)}{\sqrt{1-\Phi_s^2/c^4}}
\label{e1}
\end{equation}
where $E_0$ and $E_1$ are, respectively, total energy of the body
before and after the emission of the light as measured in $S_s$
and $c$ is speed of light in $S_s$. It should be noted that $c$
depends on the gravitational field, written as \cite{Einstein11}
\begin{equation}
c\cong c_0\Bigl(1+\frac{\Phi_s}{c^2}\Bigr)
\end{equation}
where $c_0$ is speed of light in the absence of gravitational
field, i.e. when $\Phi_s=0$. The factor multiplying $L$ in the
last term takes into account gravitation of the emitted energy $L$
\cite{Einstein11} and which reduces to $1$ in the absence of
gravitational field i.e. $\Phi_s=0$.

It should be noted that the last term can be approximated as
\begin{equation}
L\frac{\Bigl(1+\frac{\Phi_s}{c^2}\Bigr)}{\sqrt{1-\Phi_s^2/c^4}}\cong L+L\frac{\Phi_s}{c^2}
\end{equation}
when $c^2>>\Phi_s$. Einstein \cite{Einstein11} considered this
approximate expression $L+L\frac{\Phi_s}{c^2}$, representing
summation of emitted energy $L$ and gravitation of energy
$L\frac{\Phi_s}{c^2}$, for his derivation of gravitation of energy
and showed increase in gravitational mass equal to increase in
inertia mass for any body when the body absorbs radiation energy.
Here we should mention that if, instead of approximate expression
of type $L+L\frac{\Phi_s}{c^2}$, we consider more accurate
expression of type
$L\frac{\Bigl(1+\frac{\Phi_s}{c^2}\Bigr)}{\sqrt{1-\Phi_s^2/c^4}}$
in the Einstein's analysis \cite{Einstein11} on increase in
gravitational mass of the body when it absorbs $E$ amount of
radiation energy, we would obtain this increase in mass equal to
\begin{equation}
\frac{E}{\Phi_s}\Bigl[\frac{1+\frac{\Phi_s}{c^2}}{\sqrt{1-\Phi_s^2/c^4}
}-1 \Bigr]
\end{equation}
instead of $E/c^2$ as suggested by Einstein \cite{Einstein11}.

Now, the conservation of energy principle for the body $W$ as
observed from $S_v$ can be written as
\begin{equation}
H_0=H_1+\frac{L}{\sqrt{1-v^2/c^2}}\frac{\Bigl(1+\frac{\Phi_v}{c^2}\Bigr)}{\sqrt{1-\Phi_v^2/c^4}}
\label{e2}
\end{equation}
where $H_0$ and $H_1$ are, respectively, total energy of the body
before and after the emission of the light as measured in $S_v$.
While writing Eq. (\ref{e2}), we have used expression
\begin{equation}
\frac{L}{\sqrt{1-v^2/c^2}}
\end{equation}
for emitted energy as measured in $S_v$. Subtracting Eq.
(\ref{e1}) from Eq. (\ref{e2}) and using $\Phi_v=\Phi_s$ yield
\begin{equation}
(H_0-E_0)-(H_1-E_1)=L[\frac{1}{\sqrt{1-v^2/c^2}}-1]\frac{\Bigl(1+\frac{\Phi_s}{c^2}\Bigr)}
{\sqrt{1-\Phi_s^2/c^4}}. \label{eq2a}
\end{equation}
Now, the total energy of the body is summation of potential energy
$P$, kinetic energy $K$ and energy related to internal state of
the body which we refer here as internal energy $I$. We now write
total energies ($E_0, E_1, H_0$ and $H_1$) before and after the
emission in reference frames $S_s$ and $S_v$ in terms of
potential, kinetic and internal energies.

In $S_s$,
\begin{eqnarray}
E_0=M_s\Phi_s+M_sI_s \label{ne1}\\
E_1=(M_s-m_s)\Phi_s+(M_s-m_s)I_s'.
\end{eqnarray}
Here $M_s$ is mass of the stationary body $W$ before the emission,
$m_s$ is decrease in mass of the body due to the emission, $I_s$
and $I_s'$ are internal energy per unit mass of the body before
and after the emission, respectively, and all are measured in
$S_s$ in which the body is stationary all the time.

As measured in $S_v$, we denote the mass of the moving body $W$
before the emission by $M_v$, rest mass of the body before the
emission when at rest in $S_v$ by $M_v^s$, decrease in rest mass
due to the emission by $m_v^s$, potential energy per unit mass by
$\Phi_v$, internal energy per unit mass of the body before and
after the emission by $I_v$ and $I_v'$, respectively. With these
notations we can write total energy of the body before and after
the emission as measured in $S_v$ as
\begin{eqnarray}
H_0=K_0+M_v^s\Phi_v+M_v^sI_v,\label{pr1}\\
H_1=K_1+(M_v^s-m_v^s)\Phi_v+(M_v^s-m_v^s)I_v',\label{pr2}
\end{eqnarray}
where $K_0$ and $K_1$ represent kinetic energies before and after
the emission, respectively. Substituting Eqs.
(\ref{ne1})-(\ref{pr2}) into Eq. (\ref{eq2a}), we obtain
\begin{equation}
K_0-K_1=L[\frac{1}{\sqrt{1-v^2/c^2}}-1]\frac{\Bigl(1+\frac{\Phi_s}{c^2}\Bigr)}{\sqrt{1-\Phi_s^2/c^4}}
\end{equation}
where we have used the fact that $M_s=M_v^s, m_s=m_v^s,
\Phi_v=\Phi_s$ and internal energy per unit mass of the body is
same in $S_s$ and $S_v$, i.e. $I_v=I_s, I_v'=I_s'$. Further
incorporation of expressions for kinetic energies $K_0$ and $K_1$
\cite{Einstein05c} yield
\begin{equation}
m_sc^2=L\frac{\Bigl(1+\frac{\Phi_s}{c^2}\Bigr)}{\sqrt{1-\Phi_s^2/c^4}},
\label{finf}
\end{equation}
exhibiting the effect of gravitation (through $\Phi_s$) on the
decrease in mass due to the emission of energy $L$. When
$\Phi_s=0$, this final Eq. (\ref{finf}) reduces to the well known
mass-energy equivalence relation $L=m_sc_0^2$ derived by Einstein
\cite{Einstein05}.


\begin{thebibliography}{1}

\bibitem{Einstein05}
A. Einstein,  in {\em The Principle of Relativity}, edited by W. Perrett and
  G.~B. Jeffery (Dover Publications, Inc., New York, 1952), p.\ 69-71.

\bibitem{Fadner88}
W.~L. Fadner, Am. J. Phys. {\bf 56},  114  (1988); J. Stachel and
R. Torretti, Am. J. Phys. {\bf 50}, 760 (1982); {\em The EINSTEIN
Myth and the IVES papers}, edited by D. Turner and R. Hazelett
(The Devin-Adair Company, 1979)

\bibitem{Einstein11}
A. Einstein,  in {\em The Principle of Relativity}, edited by W. Perrett and
  G.~B. Jeffery (Dover Publications, Inc., New York, 1952), p.\ 99-108.

\bibitem{Einstein05c}
A. Einstein,  in {\em The Principle of Relativity}, edited by W. Perrett and
  G.~B. Jeffery (Dover Publications, Inc., New York, 1952), p.\ 37-65.

\end{thebibliography}
\end{document}